\documentclass[cits]{PoS}\usepackage{amsmath,amsbsy,cite} 

\title{Compact Flavour-Exotic Tetraquarks\\in Large-${N_{\rm c}}$
QCD --- To Be or Not to Be?}\ShortTitle{Compact Flavour-Exotic
Tetraquarks in Large-$N_{\rm c}$ QCD --- To Be or Not to Be?}
\author{Wolfgang Lucha\\Institute for High Energy Physics, Austrian
Academy of Sciences, Nikolsdorfergasse 18,\\A-1050 Vienna, Austria
\\E-mail: \email{Wolfgang.Lucha@oeaw.ac.at}}\author{\speaker{Dmitri
Melikhov}\\Institute for High Energy Physics, Austrian Academy of
Sciences, Nikolsdorfergasse 18,\\A-1050 Vienna, Austria, and\\
D.~V.~Skobeltsyn Institute of Nuclear Physics, M.~V.~Lomonosov
Moscow State University,\\119991, Moscow, Russia, and\\Faculty of
Physics, University of Vienna, Boltzmanngasse 5, A-1090 Vienna,
Austria\\E-mail: \email{dmitri\_melikhov@gmx.de}}\author{Hagop
Sazdjian\\Institut de Physique Nucl\'eaire, CNRS-IN2P3,
Universit\'e Paris-Sud, Universit\'e Paris-Saclay,\\91405 Orsay
Cedex, France\\E-mail: \email{sazdjian@ipno.in2p3.fr}}

\abstract{We analyse exotic tetraquark mesons --- bound states
formed by two quarks and two antiquarks --- for the specific case
of four different quark flavours within the framework of a
well-defined limit of quantum chromodynamics characterized by the
correlated growth without bound of the number of colour degrees of
freedom and approach to zero of the coupling constant of the
strong interactions. On the one hand, the assumption that the
tetraquarks of the kind defined above show up, as poles in the
amplitudes for the scattering of two ordinary mesons with flavour
quantum numbers that match those of these tetraquarks, already at
lowest possible order in an expansion in inverse powers of the
number of colours implies the existence of two types of
flavour-exotic tetraquarks, distinguishable by their dominant
transitions to two ordinary mesons. On the other hand, we are
aware of merely a single, unique colour-singlet arrangement of two
quarks and two antiquarks capable of tolerating a compact
flavour-exotic tetraquark, a bound state of colour-antisymmetric
diquark and antidiquark. In view of these two clearly
contradictory observations, we are led to the plausible conclusion
that, within the considered limiting case of quantum
chromodynamics, a conceivable explanation of the riddle might
consist in the non-existence of any flavour-exotic~tetraquarks in
form of narrow~states.}

\FullConference{XIII Quark Confinement and the Hadron Spectrum ---
Confinement2018\\31 July -- 6 August 2018\\Maynooth University,
Ireland}

\begin{document}\section{Identifying Tetraquark-Phile Feynman
Diagram Contributions in Large-$N_{\rm c}$ Limit}\label{I}
Recently, we embarked on a systematic study \cite{TQ1,TQ2,TQ3} of
basic qualitative properties of tetraquark mesons, $T=(\bar
q_a\,q_b\,\bar q_c\,q_d)$, regarded as bound states of two quarks
$q_b,q_d$ and two antiquarks $\bar q_a,\bar q_c$~of flavour
quantum numbers $a,b,c,d\in\{u,d,s,c,b\}$ and masses
$m_a,m_b,m_c,m_d$, respectively, by trying to utilize tetraquark
appearances in the scattering of two ordinary mesons, of momenta
$p_1$ and $p_2$, to two ordinary mesons, of momenta $p_1'$ and
$p_2'$. Suppressing all aspects of spin and parity, we adopt as
interpolating operator of a given ordinary meson $M_{\bar ab}$
some quark-bilinear current $j_{\bar ab}\equiv\bar q_a\,q_b$~whose
vacuum--meson matrix element defines the leptonic decay constant
of this meson, $f_{M_{\bar ab}}$, according~to\begin{equation}
\langle0|j_{\bar ab}|M_{\bar ab}\rangle\equiv f_{M_{\bar ab}}\ne0\
.\label{f}\end{equation}Let us confine ourselves to
\emph{compact\/} tetraquarks, tightly bound (in contrast to
molecular-type)~states.

For these investigations \cite{TQ1,TQ2,TQ3}, our \emph{tool\/} of
choice is a particular limiting case of a generalization of
quantum chromodynamics (QCD) to an \emph{arbitrary\/} number
$N_{\rm c}$ of colour degrees of freedom. QCD is a gauge theory
relying on ${\rm SU}(3)$, with all quarks in its 3-dimensional
fundamental representation. Large-$N_{\rm c}$ QCD \cite{GH,EW} is
a theory invariant under transformations of the gauge group ${\rm
SU}(N_{\rm c})$, defined by a related approach of $N_{\rm c}$ to
infinity and its strong coupling $g_{\rm s}$ or fine-structure
constant $\alpha_{\rm s}$ to~zero,$$g_{\rm s}\propto\frac{1}
{\sqrt{N_{\rm c}}}\xrightarrow[N_{\rm c}\to\infty]{}0\qquad
\Longleftrightarrow\qquad\alpha_{\rm s}\equiv\frac{g_{\rm
s}^2}{4\pi}\propto\frac{1}{N_{\rm c}}\xrightarrow[N_{\rm c}\to
\infty]{}0\ ,$$where all quarks transform according to the $N_{\rm
c}$-dimensional fundamental representation of ${\rm SU}(N_{\rm
c})$. Compared with QCD, its large-$N_{\rm c}$ limit exhibits a
considerably reduced complexity. Among others, it immediately
predicts, as the large-$N_{\rm c}$ behaviour of an ordinary-meson
leptonic decay constant~\cite{EW},$$f_{M_{\bar ab}}\propto
\sqrt{N_{\rm c}}\qquad\mbox{for}\quad N_{\rm c}\to\infty\
.$$Instigated by Ref.~\cite{SW}, various aspects of tetraquarks
have been discussed along similar lines \cite{KP,CL,MPR1,MPR2}.

Our actual \emph{targets\/} for the application of the limit
$N_{\rm c}\to\infty$ (and a $1/N_{\rm c}$ expansion thereabout)
are appropriate four-current Green functions. From these, we
derive the amplitudes for the scattering of two ordinary mesons
into two ordinary mesons, both pairs of mesons
carrying,~of~course,~the~flavour quantum numbers of any tetraquark
in the focus of our interest. These~scattering amplitudes~we~then
inspect for the presence of a pole betraying the existence of a
compact-tetraquark intermediate state.

From our point of view, our first \emph{task\/} in this enterprise
is the formulation of rigorous criteria for the selection of those
Feynman diagrams that offer the perspective of exhibiting
singularities related to four-quark intermediate states that
eventually can contribute to the formation of a tetraquark pole.
Feynman diagrams conforming to this requirement are called
tetraquark-phile \cite{ph1,ph2,ph3,ph4}. We propose two necessary,
but not sufficient, consistency criteria expressed in terms of the
Mandelstam variable$$s\equiv(p_1+p_2)^2=(p_1'+p_2')^2\ .$$
\begin{enumerate}\item A \emph{tetraquark-phile\/} Feynman diagram
must depend \emph{nontrivially}, \emph{viz.},
\emph{non-polynomially}, on $s$.\item A \emph{tetraquark-phile\/}
Feynman diagram must support an adequate four-quark intermediate
state and develop a corresponding branch cut, starting at a branch
point\footnote{For each Feynman diagram under consideration, the
existence or non-existence of such a singularity can be decided
straightforwardly by means of the Landau equations \cite{LDL}.
See, \emph{e.g.}, Ref.~\cite[App.~A]{TQ2} for a variety of
illustrative examples.} $\hat s=(m_a+m_b+m_c+m_d)^2$.
\end{enumerate}

\section{\emph{Flavour-Exotic\/} Tetraquarks: Pairwise Appearance
in Ordinary-Meson Scattering}Next, we should specify the
quark-flavour composition of the compact tetraquarks we set~out~to
study. A list of conceivable flavour combinations in tetraquarks
can be found in Table~1 of~Ref.~\cite{ph4}. Here, we adhere to the
genuinely exotic case of the flavours of all four (anti-)quarks
being disparate.

\subsection{Two Categories of Scattering Processes Exhibiting
Unlike Behaviour in the Limit $N_{\rm c}\to\infty$}
\begin{figure}[hbt]\centering
\includegraphics[scale=.52909,clip]{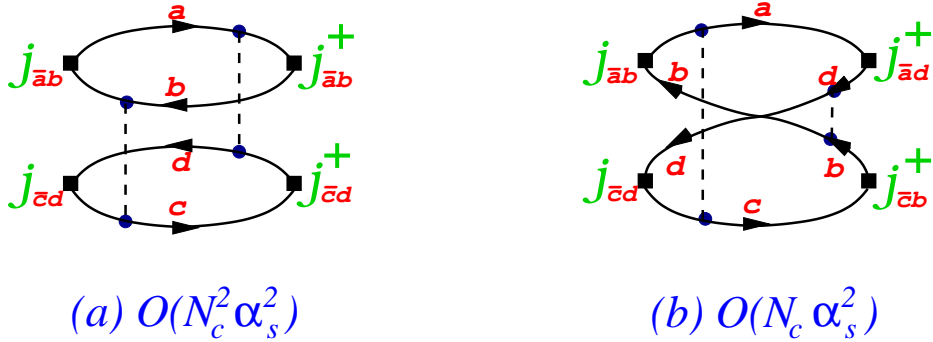}
\caption{Typical representatives of $N_{\rm c}$-leading
tetraquark-phile Feynman diagrams potentially developing a
flavour-exotic tetraquark pole in either flavour-preserving (a) or
flavour-reshuffling (b) scattering amplitudes.}\label{f12}
\end{figure}

The possible
end-product of any scattering of two mesons $M_{\bar ab}$ and
$M_{\bar cd}$ to two ordinary mesons might be either the same two
mesons, $M_{\bar ab}$ and $M_{\bar cd}$, or two different mesons,
$M_{\bar ad}$ and $M_{\bar cb}$, resulting from a redistribution
of the available quark flavours. Within the envisaged quest for
tetraquark poles in scattering amplitudes, we hence have to take
into account two disjoint classes of Green functions:
\begin{subequations}\begin{align}&\mbox{flavour-preserving
correlators}&&\quad\langle T(j_{\bar ab}\,j_{\bar
cd}\,j^\dag_{\bar ab}\,j^\dag_{\bar cd})\rangle\ ,\qquad\langle
T(j_{\bar ad}\,j_{\bar cb}\,j^\dag_{\bar ad}\,j^\dag_{\bar
cb})\rangle\ ;\qquad\label{ep}\\&\mbox{flavour-reshuffling
correlators}&&\quad\langle T(j_{\bar ad}\,j_{\bar
cb}\,j^\dag_{\bar ab}\,j^\dag_{\bar cd})\rangle\ .\label{er}
\end{align}\end{subequations}The actual large-$N_{\rm c}$
dependence of the tetraquark-phile contributions (identified by a
subscript T) to these Green functions at leading order in their
series expansion in powers of $1/N_{\rm c}$ \cite{TQ1} can be
read~off from (of course, $N_{\rm c}$-subleading) Feynman diagrams
of the type illustrated by the examples~in Fig.~\ref{f12}.

From the topological point of view, the discussion of
tetraquark-phile correlators is presumably facilitated if
imagining any encountered Feynman diagram as residing on a
cylinder \cite{TQ3} with surface spanned or even bordered by quark
lines. Let us now look at our two classes of scattering processes.

\subsection{Flavour-Preserving Scattering at Leading
Tetraquark-Phile Order of its $1/N_{\rm c}$ Expansion}\label{P}
\begin{figure}[hbt]\centering
\includegraphics[scale=.52909,clip]{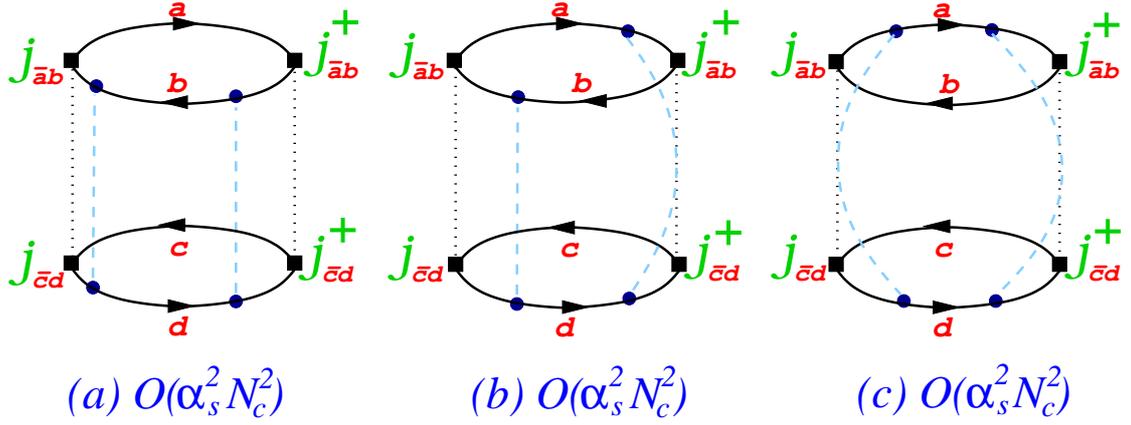}\caption{Cylinder
interpretation of the flavour-retaining $N_{\rm c}$-leading
tetraquark-phile Feynman diagrams \cite{TQ3}. Dashed blue lines:
planar gluons (of least tetraquark-phile number two). Dotted black
lines: cylinder surface.}\label{fp}\end{figure}

\begin{figure}[htb]\centering
\includegraphics[scale=.52909,clip]{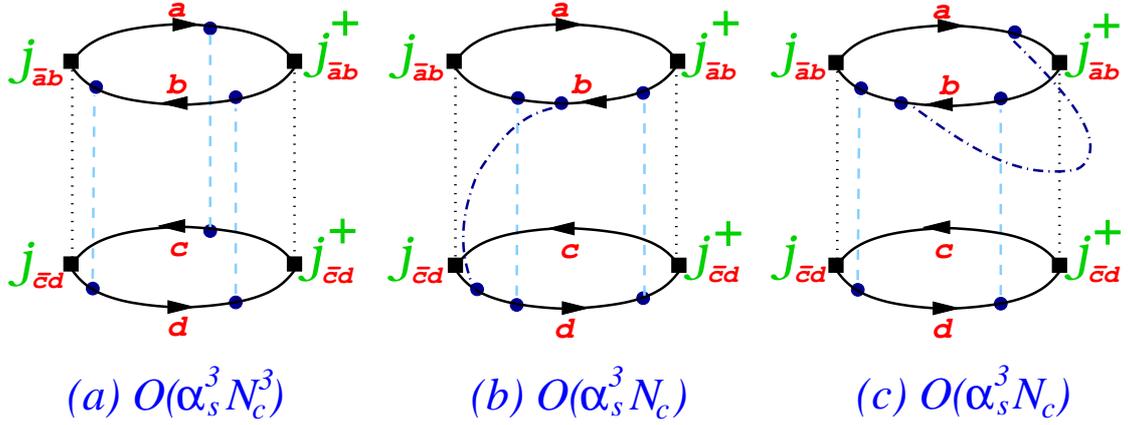}\caption{Amending
Fig.~\protect\ref{fp}(a) by one further (a) planar gluon (dashed)
or (b,c) nonplanar gluon (dot-dashed).}\label{fp+}\end{figure}

In
the flavour-preserving channel, the $N_{\rm c}$-leading Feynman
diagrams (Fig.~\ref{fp}) involve two gluon exchanges, \emph{i.e.},
four quark--gluon vertices, equivalent to perturbative order
$\alpha_{\rm s}^2$, and two colour loops (contributing a factor
$N_{\rm c}^2$): the resulting $N_{\rm c}$ dependence of such a
contribution is $O(\alpha_{\rm s}^2\,N_{\rm c}^2)=O(N_{\rm c}^0)$.
The addition of a further gluon amounts to two additional
quark--gluon vertices, raising the power of $\alpha_{\rm s}$ by
one: embedding this gluon in the cylinder surface increases the
number of colour loops~by one (providing a further factor $N_{\rm
c}$) and implies the earlier $N_{\rm c}$ dependence $O(\alpha_{\rm
s}^3\,N_{\rm c}^3)=O(N_{\rm c}^0)$ [Fig.~\ref{fp+}(a)]; if,
however, the arrangement is such that this gluon no longer fits to
the cylinder surface, the number of colour loops gets
\emph{reduced\/} by one, and likewise the $N_{\rm c}$ order to
$O(\alpha_{\rm s}^3\,N_{\rm c})=O(N_{\rm c}^{-2})$
[Figs.~\ref{fp+}(b,c)].\pagebreak

\begin{figure}[htb]\centering
\includegraphics[scale=.39768,clip]{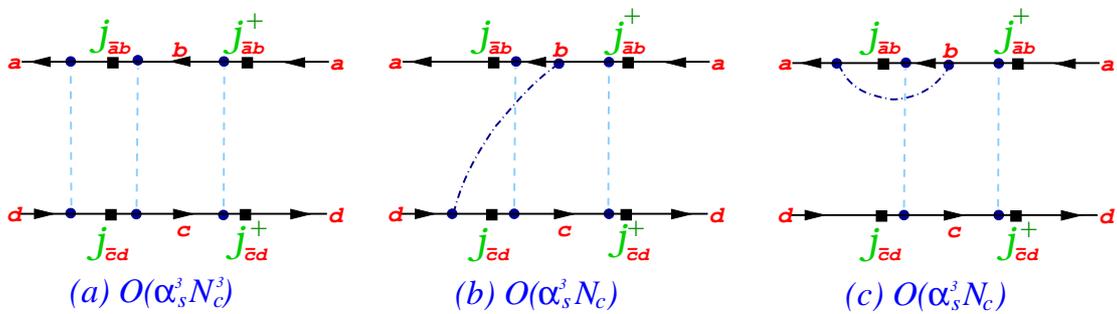}\caption{Planar
interpretation of the Feynman diagrams of Fig.~\protect\ref{fp+},
demanding identification of points at~$\pm\infty$.}\label{fp+f}
\end{figure}

Unfolding the imagined cylinder surfaces to planes
(Fig.~\ref{fp+f}), by breaking up the two quark~loops in the
Feynman diagrams of Fig.~\ref{fp+}, explains the distinction of
the notions planar or nonplanar gluon.

Since we have to deal with the two classes of available scattering
channels (\ref{ep}) and (\ref{er}), we take the liberty of
foreseeing notationally, already from the very beginning, the
possible existence of two different tetraquarks $T_A$ and $T_B$ of
the same exotic flavour content, with masses $m_{T_A}$ and
$m_{T_B}$, and respective large-$N_{\rm c}$ preferred coupling to
one of the ordinary-meson pairs $M_{\bar ab}+M_{\bar cd}$ and
$M_{\bar ad}+M_{\bar cb}$. For the \emph{$N_{\rm c}$-leading
tetraquark-phile\/} contributions to flavour-preserving Green
functions, in terms~of generic decay constants $f_M$ and
tetraquark--two-ordinary-meson transition amplitudes $A$ we then
get\begin{align}\langle T(j_{\bar ab}\,j_{\bar cd}\,j^\dag_{\bar
ab}\,j^\dag_{\bar cd})\rangle_{\rm T}&=f_M^4\left(\frac{|A(M_{\bar
ab}\,M_{\bar cd}\leftrightarrow T_A)|^2}{p^2-m^2_{T_A}}+
\frac{|A(M_{\bar ab}\,M_{\bar cd}\leftrightarrow T_B)|^2}
{p^2-m^2_{T_B}}\right)+\cdots=O(N_{\rm c}^0)\ ,\label{fp1}\\[1ex]
\langle T(j_{\bar ad}\,j_{\bar cb}\,j^\dag_{\bar ad}\,j^\dag_{\bar
cb})\rangle_{\rm T}&=f_M^4\left(\frac{|A(M_{\bar ad}\,M_{\bar
cb}\leftrightarrow T_A)|^2}{p^2-m^2_{T_A}}+\frac{|A(M_{\bar
ad}\,M_{\bar cb}\leftrightarrow T_B)|^2}{p^2-m^2_{T_B}}\right)
+\cdots=O(N_{\rm c}^0)\ .\label{fp2}\end{align}

\subsection{Flavour-Regrouping Scattering at Lowest
Tetraquark-Phile Order in its $1/N_{\rm c}$~Expansion}\label{R}
\begin{figure}[htb]\centering
\includegraphics[scale=.52909,clip]{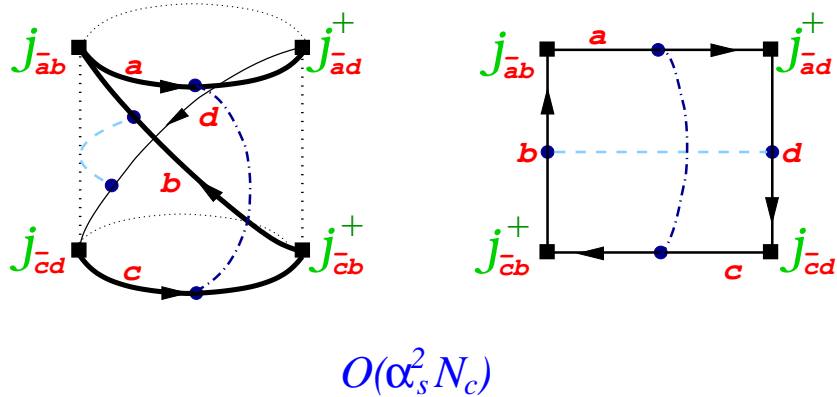}
\caption{Cylindric (left) and unfolded (right) portrayal of a
flavour-rearranging \emph{$N_{\rm c}$-leading tetraquark-phile\/}
Feynman diagram \cite{TQ3}, involving a planar and a nonplanar
gluon distinguished by dashed vs.\ dot-dashed lines.}\label{fr}
\end{figure}

In
contrast to the above flavour-retaining case, in the
flavour-reshuffling channel the $N_{\rm c}$-leading Feynman
diagrams (Fig.~\ref{fr}) need one nonplanar gluon exchange.
Tetraquark-phile planar options do not exist. The $N_{\rm
c}$-leading tetraquark-phile contributions to flavour-rearranging
Green functions~read\begin{align}\langle T(j_{\bar ad}\,j_{\bar
cb}\,j^\dag_{\bar ab}\,j^\dag_{\bar cd})\rangle_{\rm T}&=f_M^4
\left(\frac{A(M_{\bar ab}\,M_{\bar cd}\leftrightarrow T_A)\,
A(T_A\leftrightarrow M_{\bar ad}\,M_{\bar cb})}{p^2-m^2_{T_A}}
\right.\nonumber\\ &\hspace{4.65ex}+\left.\frac{A(M_{\bar
ab}\,M_{\bar cd}\leftrightarrow T_B)\,A(T_B\leftrightarrow M_{\bar
ad}\,M_{\bar cb})}{p^2-m^2_{T_B}}\right)+\cdots=O(N_{\rm c}^{-1})\
.\label{frT}\end{align}

\subsection{Constraining the Spectra of Tetraquarks Formed at
$N_{\rm c}$-Leading Tetraquark-Phile Order}Assuming the tetraquark
masses $m_{T_{A,B}}$ not to grow with $N_{\rm c}$ without bound
but to remain finite~in the limit $N_{\rm c}\to\infty$, let us now
explore some implications for the resulting tetraquark spectra.
Lacking convincing arguments why tetraquarks cannot emerge at the
largest $N_{\rm c}$ order tolerated by the criteria of
Sect.~\ref{I}, we expect them to reveal their presence indeed at
the leading tetraquark-phile order of~$N_{\rm c}$.

The most self-evident starting point in the attempt to satisfy the
large-$N_{\rm c}$ behaviour derived,~for flavour-exotic
tetraquarks, in Sects.~\ref{P} and \ref{R} is to assume the
existence of only a single tetraquark state $T$. However, upon
setting $T_A=T_B=T,$ Eqs.~(\ref{fp1}), (\ref{fp2}), and
(\ref{frT}) collapse to the requirements\begin{align*}&
A(T\leftrightarrow M_{\bar ab}\,M_{\bar cd})=O(N_{\rm c}^{-1})\
,\qquad A(T\leftrightarrow M_{\bar ad}\,M_{\bar cb})=O(N_{\rm
c}^{-1})\ ,\\[1ex]&A(T\leftrightarrow M_{\bar ab}\,M_{\bar
cd})\,A(T\leftrightarrow M_{\bar ad}\,M_{\bar cb})=O(N_{\rm
c}^{-3})\end{align*} on the three involved transition amplitudes
$A$: the prevailing mutual contradiction is pretty
evident.\footnote{Trivially, the addition of a single quark loop
to a gluon line modifies the $N_{\rm c}$ dependence of a given
Feynman~diagram by one unit. This observation may be abused in
order to bring the large-$N_{\rm c}$ scaling of these two classes
of tetraquark-phile contributions into agreement. However, such
ad-hoc modification renders any conclusion drawn from inspection
of some large-$N_{\rm c}$ behaviour meaningless. We consequently
disregard such unmotivated way of adjusting one's desired~$N_{\rm
c}$~scaling.}

In contrast, allowing for two tetraquarks $T_{A,B}$, of same
flavour content but different couplings~to ordinary mesons,
enables us to offer a solution to the requests implied by
Eqs.~(\ref{fp1}), (\ref{fp2}), and (\ref{frT}):\begin{align*}
\underbrace{A(T_A\leftrightarrow M_{\bar ab}\,M_{\bar cd})
=O(N_{\rm c}^{-1})}_{\mbox{$\Longrightarrow\qquad\Gamma(T_A)
=O(N_{\rm c}^{-2})$}}\qquad\stackrel{N_{\rm c}}{>}\qquad
A(T_A\leftrightarrow M_{\bar ad}\,M_{\bar cb})=O(N_{\rm c}^{-2})\
,&\\[1ex]A(T_B\leftrightarrow M_{\bar ab}\,M_{\bar cd})=O(N_{\rm
c}^{-2})\qquad\stackrel{N_{\rm c}}{<}\qquad\underbrace{A(T_B
\leftrightarrow M_{\bar ad}\,M_{\bar cb})=O(N_{\rm c}^{-1})}_
{\mbox{$\Longrightarrow\qquad\Gamma(T_B)=O(N_{\rm c}^{-2})$}}\ .&
\end{align*}Accordingly, under the various assumptions mentioned
above we are \emph{unavoidably\/} led to consider~as rather
probable the existence of (at least) two --- more precisely,
pairwise existence of --- differently decaying tetraquarks, $T_A$
and $T_B$, of genuinely exotic quark-flavour composition. Their
$N_{\rm c}$-dominant decay channels control the identical
large-$N_{\rm c}$ behaviour of their total decay widths
$\Gamma(T_A)$ and~$\Gamma(T_B)$,$$\Gamma(T_A)=O(N_{\rm c}^{-2})\
,\qquad\Gamma(T_B)=O(N_{\rm c}^{-2})\ ,$$and, consequently, the
nature of these flavour-exotic tetraquark states: they are both
narrow~mesons.

\section{\emph{Compact\/} Flavour-Exotic Tetraquarks: Formation
Mechanisms vs.\ Large-$N_{\rm c}$ QCD}With respect to the colour
configuration inside the generic genuinely flavour-exotic
tetraquarks $T=(\bar q_a\,q_b\,\bar q_c\,q_d)$, we see two
possibilities for the two-step formation of any such
colour-singlet~state from two quarks and two antiquarks in the
($N_{\rm c}$-dimensional) fundamental representation of ${\rm
SU}(N_{\rm c})$: The formation of two colour-singlet
\emph{quark--antiquark\/} states, followed by the formation of a
(loosely bound) molecular-type tetraquark, for which there are two
options $(\bar q_a\,q_b)\,(\bar q_c\,q_d)$ and $(\bar
q_a\,q_d)\,(\bar q_c\,q_b)$, or the formation of a diquark and an
antidiquark, followed by the formation of a compact tetraquark,
for which there is, however, only a single option in the genuinely
flavour-exotic case, $(\bar q_a\,\bar q_c)\,(q_b\,q_d)$, as the
latter mechanism necessitates the (anti-)diquarks to transform
according to the antisymmetric $N_{\rm c}\,(N_{\rm
c}-1)/2$-dimensional representation of ${\rm SU}(N_{\rm c})$. We
are hence confronted with two conflicting findings: On the one
hand, the inspection of scattering amplitudes at $N_{\rm
c}$-leading order \cite{TQ1,TQ2} suggests the \emph{pairwise\/}
existence of flavour-exotic tetraquarks, distinguishable by their
preferred decay modes. On the other hand, among the described
two-phase creation processes of colour-singlet bound~states of two
quarks and two antiquarks there is just \emph{one\/} promising
candidate for the formation~of~\emph{compact\/} tetraquarks. A
solution to this riddle may be the \emph{nonexistence\/} of
\emph{compact flavour-exotic\/} tetraquarks.\pagebreak

\noindent{\bf Acknowledgement}. D.~M.\ is supported by the
Austrian Science Fund (FWF), Project P29028-N27.


\begin{thebibliography}{99}
\bibitem{TQ1}W.~Lucha, D.~Melikhov, and H.~Sazdjian, Phys.~Rev.~D
{\bf 96} (2017) 014022, arXiv:1706.06003 [hep-ph].
\bibitem{TQ2}W.~Lucha, D.~Melikhov, and H.~Sazdjian,
Eur.~Phys.~J.~C {\bf 77} (2017) 866, arXiv:1710.08316 [hep-ph].
\bibitem{TQ3}W.~Lucha, D.~Melikhov, and H.~Sazdjian, Phys.~Rev.~D
{\bf 98} (2018) 094011, arXiv:1810.09986 [hep-ph].
\bibitem{GH}G.~'t Hooft, Nucl.~Phys.~B \textbf{72} (1974) 461.
\bibitem{EW}E.~Witten, Nucl.~Phys.~B \textbf{160} (1979) 57.
\bibitem{SW}S.~Weinberg, Phys.~Rev.~Lett.~{\bf 110} (2013) 261601,
arXiv:1303.0342 [hep-ph].
\bibitem{KP}M.~Knecht and S.~Peris, Phys.~Rev.~D \textbf{88} (2013)
036016, arXiv:1307.1273 [hep-ph].
\bibitem{CL}T.~D.~Cohen and R.~F.~Lebed, Phys.~Rev.~D \textbf{90}
(2014) 016001, arXiv:1403.8090 [hep-ph].
\bibitem{MPR1}L.~Maiani, A.~D.~Polosa, and V.~Riquer, J.~High Energy
Phys.~06 (2016) 160, arXiv:1605.04839 [hep-ph].
\bibitem{MPR2}L.~Maiani, A.~D.~Polosa, and V.~Riquer, Phys.~Rev.~D
\textbf{98} (2018) 054023, arXiv:1803.06883 [hep-ph].
\bibitem{ph1}W.~Lucha, D.~Melikhov, and H.~Sazdjian, PoS
(EPS-HEP2017) 390, arXiv:1709.02132 [hep-ph].
\bibitem{ph2}W.~Lucha, D.~Melikhov, and H.~Sazdjian, PoS (Hadron
2017) 233, arXiv:1711.03925 [hep-ph].
\bibitem{ph3}W.~Lucha, D.~Melikhov, and H.~Sazdjian, EPJ Web
Conf.~{\bf 191} (2018) 04003, arXiv:1807.09630 [hep-ph].
\bibitem{ph4}W.~Lucha, D.~Melikhov, and H.~Sazdjian, EPJ Web
Conf.~{\bf 192} (2018) 00044, arXiv:1808.05519 [hep-ph].
\bibitem{LDL}L.~D.~Landau, Nucl.~Phys.~{\bf 13} (1959) 181.
\end{thebibliography}
\end{document}